\documentclass[a4paper,11pt]{article}
\usepackage{enumerate}
\usepackage{url}

\usepackage[utf8]{inputenc}

\usepackage{amsmath}
\usepackage{multirow}
\usepackage{amssymb}

\newtheorem{thm}{Theorem} 
\newtheorem{prop}[thm]{Proposition} 
 
\newtheorem{lemma}[thm]{Lemma} 
\newtheorem{theorem}[thm]{Theorem}

\newtheorem{de}[thm]{Definition}
\newtheorem{defn}[thm]{Definition}

\newcommand{\beq}{\begin{eqnarray}}
\newcommand{\eeq}{\end{eqnarray}}
\newcommand{\V}{\mathbb V}

\newcommand{\Z}{\mathbb Z}
\newcommand{\N}{\mathbb N}

\title{${\cal J}_2$ radical in automata nearrings}

\author{Tim Boykett,\\ Time's Up Research, Industriezeile 33b, 4020 Linz, Austria \\and Institute for Algebra, Johannes Kepler University, Linz, Austria\\
email:tim@timesup.org, tim.boykett@jku.at\\
and\\
Gerhard Wendt,\\ Institute for Algebra, Johannes Kepler University, Linz, Austria\\
email:gerhard.wendt@jku.at}

\begin{document}

\maketitle

\vspace{1mm}
\begin{center}
 \emph{Dedicated to Professor G\"unter Pilz on the occasion of his retirement.}

\end{center}
\vspace{2mm}

\begin{abstract}
 Looking at the automata defined over a group alphabet as a nearring, we see that they are a highly complicated structure. 
 As with ring theory, one method to deal with complexity is to look at semisimplicity modulo radical structures. 
 We find some bounds on the Jacobson 2-radical and show that in certain groups, 
 this radical can be explicitly found and the semisimple image determined.
\end{abstract}


\section{Introduction}
Automata defined over a group alphabet can be interpreted as mappings from a group $G$ into itself. 
A natural algebraic structure to study mappings from a group into itself is a nearring. 
The link between automata we are studying and nearrings will be further explained in the following. First we fix and introduce some notation and elementary facts concerning nearrings.

A \emph{nearring} $(N,+,*)$ is an algebra with two binary operations such that $(N,+)$ is a group and $(N,*)$ is a semigroup 
and the right distributive law $(a+b)*c=a*c+b*c$ holds for all $a,b,c \in N$. Usually we will omit the operation symbol
$*$ in the following. The notation we use is that of \cite{pilzbook}.

Let $0$ be the neutral element of the group $(N,+)$ of the nearring $N$. A nearring $N$ is called \emph{zero symmetric} 
if for all $n \in N$, $n*0=0$.  Nearrings arise naturally when studying mappings from 
a group $G$ into itself. Let $M(G):=\{f: G \rightarrow G |\text{f is a function}\}$. Then, $M(G)$ is a nearring with respect to 
pointwise addition of functions and function composition. In fact, any nearring can be embedded as a subnearring into some $M(G)$
for a suitable group $G$. If the zero $0$ of the group $G$ is preserved by the mappings, then we get the zero symmetric 
subnearring $M_0(G):=\{f: G \rightarrow G|\text{f is a function and} f(0)=0\}$ of $M(G)$. 
We will call a function $f: G \rightarrow G$ \emph{zero preserving} if $f(0)=0$.

A \emph{left ideal} $L$ of a nearring $N$ is a normal subgroup of the group $(N, +)$ such 
that $\forall n,m \in N\, \forall l \in L: n(m + l)-nm \in L$. 
An \emph{ideal} $I$ of a nearring is a left ideal which is 
closed under multiplication from the right hand side, so $IN \subseteq I$.
Ideals are kernels of nearring homomorphisms.
An \emph{$N$-subgroup} $S$ of the nearring $N$ is a subgroup of $(N,+)$ such that $NS \subseteq S$. In case $N$ is a zero 
symmetric nearring, a left ideal $L$ of $N$ is also an $N$-subgroup of $N$.
If $N$ is a nearring with identity element $1$, then an element $n \in N$ is called 
\emph{quasiregular} if there exists an element $m \in N$ such that $m(1-n)=1$.

In many situations it is interesting
to study the action of a nearring on a group $G$. Let $N$ be a nearring and $G$ be a group. 
A group $G$ is called an $N$-group if there is an action $N \times G \rightarrow G$, written $ng$, such that $n(mg)=(nm) g$ and $(n+m) g=ng + mg$ for all $n,m \in N$ and $g \in G$. 
A subgroup $S$ of an $N$-group $G$ will be called an $N$-subgroup of $G$ if $NS \subseteq S$.  For an $N$-group $G$ we define 
$(0:G):=\{n \in N| ng=0 \forall g\in G\}$ to be the \emph{annihilator} of this $N$-group. Annihilators of $N$-groups are always ideals in the nearring $N$. 

Let $S$ be a set of symbols. 
As is usual, $S^*$ is the set of finite sequences of symbols from the set $S$, 
with $\lambda$ being the empty sequence. 
$S^*$ forms a monoid under concatenation with $\lambda$ being the identity.
We will write this operation by juxtaposition.

We write $S^\N$ for the set of infinite sequences over $S$.

\section{Prefix preserving maps}

Let $(G,+)$ be a group so $(G^\N,+)$ is a group.
Thus
$M_0(G^\N)$ is a nearring.

\begin{de}
$n \in M_0(G^\N)$ is \emph{prefix preserving} if 
\[\forall k \in \N, x,y \in G^\N:\: x_i=y_i \forall i<k 
\Rightarrow (nx)_i = (ny)_i \forall i<k.\] 
\end{de}

Prefix preserving maps form a subnearring of $M_0(G^\N)$, we will call it $PP(G)$.
Prefix preserving maps in $M(G)$, which are not necessarily zero symmetric, will be called $PP_c(G)$

As an example, take some $f \in M_0(G)$ and $x \in G^\N$. 
Define $\bar f \in PP(G)$ by $(\bar f x)_i = f(x_i)$. This is a simple prefix preserving map.

A map in $PP(G)$ can be seen as an element of $M_0(G^n)$ for any finite $n$.
Let $x \in G^n$, $m\in PP(G)$.
Then $mx = (m(x_1,\ldots,x_n,0,0,\ldots))\vert_{\{1,\ldots,n\}}$.
We will call this action the \emph{restricted action} of $PP(G)$ on $M_0(G^n)$.

Given a group $(G,+)$, it is possible to define the nearring of 
state automata or state machines over the alphabet $G$ \cite{pilzautomata}.
If we fix the input-output alphabet $G$, a \emph{state automaton} is defined as $(Q,t,f,s)$, where
\begin{itemize}
 \item $Q$ is a set of states,
 \item $t:Q\times G \rightarrow Q$ is a state transition map,
 \item $f:Q \times G \rightarrow G$ is the output map and
 \item $s \in Q$ is the start state.
\end{itemize}

A state automaton or state machine can be seen as a mapping of $G$-sequences to $G$-sequences, which is how we will see them.
In order to calculate the mapping, we proceed as follows. 
Let $a= (Q,t,f,s)$ be a state machine, $x \in G^\N$ be the input sequence. 
Let $q_1=s$ and $q_{i+1}= t(q_i,x_i)$ for $i \geq 1$, the \emph{state sequence}. 
Then $y_i = f(q_i,x_i)$ and $y=ax \in G^\N$ is the \emph{output sequence},
the image of $x$ under the state machine mapping. 
Note that this is defined on finite as well as infinite sequences.

The concatenation of state machines and the addition of state machines are then precisely defined as composition and addition of the maps on $G^\N$.
These operations can be defined as operation on the state machines, see \cite[paragraphs before Prop 2]{pilzautomata} for details.
If two state machines agree as maps on  $G^\N$, we regard them as equal.
Call the collection of state machines on a group $SM(G)$.

We will have occasion to look at the \emph{state output maps} of a state machine $(Q,t,f,s)$. These
are the mappings $f_q:G\rightarrow G$ $g \mapsto f(q,g)$.

Note that we use the formulation of state machines as \emph{Mealy Machines}. 
If the state output map does not depend upon the input, but only upon the state, i.e.\ the mappings $f_q: g \mapsto f(q,g)$ are constant maps, then we have a \emph{Moore Machine}  (see e.g.\ \cite[p. 58]{leeseshia}).

As an example, let $f \in M_0(G)$, $Q=\{s\}$ a single element set, $\bar f (s,g)=f(g)$ and $t(s,g)=s$. 
Then $(Q,t,\bar f,s)$ is a  state machine with a single state, with state output map $\bar f_s = f$.

\begin{prop}
  Let $G$ be a group. $PP_c(G)$ is isomorphic to $SM(G)$.
\end{prop}
Proof: We have defined state machines $SM(G)$ and prefix preserving maps $PP_c(G)$ as subnearrings of $M(G^\N)$, so we need only show that they are the same as subsets.

From the definition of the sequence mapping, state machines are prefix preserving, so $SM(G) \subseteq PP_c(G)$.

 Let $n\in PP_c(G)$.
 Define $Q = G^*$.
 Define $t(q,g) = qg \in G^*$ by concatenation and $f(q,g)= n(qg)_{i+1}$ where $i$ is the length of the string $q$, $n$ acting by the restricted action on $G^{i+1}$.
Then the state machine $(Q,t,f,\lambda)$ will induce the same mapping as $n$, so $PP_c(G) \subseteq SM(G)$ and we are done.
\hfill$\Box$

We will use this equivalence often, in order to define, manipulate and analyse maps in the clearest way possible.
We will concentrate upon $PP(G)$ and are thus interested in the zero symmetric state machines in $SM(G)$.

A state $r \in Q$ is called \emph{reachable} if there is an input sequence $x \in G^\N$ such that
$r=q_i$ for some $q_i$ in the state sequence $q$. 
It is clear that unreachable states do not affect the properties of state machines.
A state $r\in Q$ is \emph{0-reachable} if for some $i$, $r=q_i$ in the state sequence induced by the zero sequence $(0,0,\ldots)$.
\begin{lemma}
 A state machine is zero symmetric if the state output maps $f_q$ are 0-preserving for every state that is 0-reachable.
\end{lemma}
Proof: Let $a=(Q,t,f,s)$ be a zero symmetric state machine. We know that a state machine is prefix preserving. 
Let $r\in Q$ be 0-reachable, $r=q_i$ with $q$ being the state sequence from the zero input sequence. 
Thus $f(q,0)=0$ and we see that every 0-reachable state has a 0-preserving state output  map.

Let $a=(Q,t,f,s)$ be a state machine with all 0-reachable states having 0-preserving state output maps.
Then $a(0,0,\ldots) = (0,0,\ldots)$ so $a$ is 0-preserving and hence $a$ is a zero symmetric state machine.
\hfill$\Box$

The following result indicates that the nearring of prefix preserving maps is complex and complicated.
Let $\V(G)$ be the variety generated by the group $G$.

\begin{theorem}
 Let $K$ be a finite group. Then for all finite groups $G \in \V(K)$, $M_0(G) < PP(K)$.
\end{theorem}

Proof:
We will show that we can encode $G$ into $K^\N$ and mappings in $M_0(G)$ as state automata mappings on $K$.

From \cite[Thm 10.16]{burrsan}, $G$ is a homomorphic image of a subgroup $S$ of a finite power of $K$, so there exists
some natural number $n$ and some $\beta:S \subseteq K^n \rightarrow G$ which is a homomorphism. 
Let $\alpha:G \rightarrow S$ be some mapping such that $\beta \circ \alpha$ is the identity on $S$.

Let $f \in M_0(G)$ be arbitrary. Define $F:S\rightarrow S$ by $F=\alpha f\beta$.

Define $Q = \cup_{j=0,\ldots, n-1} K^j \times S \times \{1,\ldots,n\} \cup \{z\}$.
$Q$ is thus made up of the cartesian product of $K$-sequences up to length $n-1$, 
a recognized element of $S$ and an index. 
The state $z$ indicates an \emph{error} state, which will not be reached by correctly encoded inputs.
Let the initial state be $s=(\lambda,0,1)$.
Define the state transition function $t:Q\times K \rightarrow Q$ as
\begin{itemize}
 \item $t((q_1,0,1),k) = (q_1k,q_1k,1)$ if $q_1k \in S$.
 \item $t((q_1,0,1),k) = (q_1k,0,1)$ if $q_1k$ is a prefix of some element of $S$.
 \item $t((q_1,0,1),k) = z$ otherwise.
 \item $t((q_1,s,n),k) = (q_1k,s,1)$ if $q_1k =s$.
 \item $t((q_1,s,i),k) = (q_1k,s,1+1)$ if $q_1k$ is a prefix of $s$.
 \item $t((q_1,s,i),k) = z$ otherwise.
 \item $t(z,k)=z$
\end{itemize}
The state transition function recognises an input sequence as consisting of a sequence of $0$s of length a multiple of $n$, followed by repeats of an element of $S$.

Define the output function $o:Q\times K \rightarrow K$ as
\begin{itemize}
 \item $o((q_1,0,1),k)=0$
 \item $o((q_1,s,i),k)=F(s)_i$ for $s\in S$
 \item $o(z,k)=0$
\end{itemize}
The output map implements componentwise the mapping $F$ on the recognized element of $S$.

For the following, let $\hat i = ((i-1) \mod n)+1$.

The encoding map $e:G\rightarrow K^\N$ has $e(g)_i = \alpha(g)_{\hat i}$. 
The decoding map $d:K^\N \rightarrow G$ is more complex.
If there exists some $m\in \N$ such that $\forall i \geq mn$, $x_i = x_{i-n}$, i.e.\ $x$ cycles, then
$d(x) = \beta(x_{mn+1},x_{mn+2},\ldots,x_{mn+n})$, otherwise $d(x)=0$.

Then the automaton defined as $a=(Q,t,o,\lambda)$ on $K$ acts on $G$ by $ag := d\circ a \circ e (g)$. 
We claim that $ag = f(g)$.

Let $x \in K^\N$ be of the form $x=(0,\ldots,0,s_1,\ldots,s_n,s_1,\ldots)$ with $mn$ $0$s in the prefix, 
$(s_1,\ldots,s_n)=\alpha(g) \in G$. 
Note that $e(g)$ is of this form with $m=0$.
Then $a(x)$ will have the state sequence $q$ with
$q_i = (o_i,0,1)$ with $o_i$ being a sequence of $(i \mod n)$ $0$s, for $i \leq mn$, 
$q_i= ((s_1,\ldots,s_{\hat i}),0,1)$ for $mn < i < (m+1)n$, 
$q_{(m+1)n} = ((s_1,\ldots,s_n),(s_1,\ldots,s_n),1)$
and
$q_i=((s_1,\ldots,s_{\hat i}),(s_1,\ldots,s_n),\hat i)$ for $i > (m+1)n$.

This will give the output sequence $y=ax \in K^\N$ with 
$y_i = 0$ for $i \leq (m+1)n$ and 
$y_i= (F((s_1,\ldots,s_n)))_{\hat i}$ for $i > (m+1)n$.
Let $g \in G$, $x=e(g)$.
Then $d(y) = \beta(F(s_1,\ldots,s_n)) = \beta \circ \alpha \circ f \circ \beta \circ \alpha (g) = f(g)$.

We have shown that the above construction maps $M_0(G)$ injectively into $PP(K)$.

Let $f_1,f_2 \in M_0(G)$, $a_1,a_2 \in PP(K)$ their encodings by the above construction.
Then $d (a_1 a_2) e=(d a_1 e)(d a_2 e)$ by the way $d$ ignores initial $0$ sequences, 
so we see that the mapping of $M_0(G)$ into $PP(K)$ is a multiplicative homomorphism. Similarly it is an additive homomorphism. 
Thus the construction is an isomorphic embedding and we are done. 
\hfill$\Box$

Note that the complex definition of $d$ was necessary because the automaton delays its output by $n$ places. 
Thus the composition of two such automata will delay by $2n$ places.

One approach to deal with complex structures in ring and nearring theory in to consider semisimple rings and nearrings.
Our goal in this paper is to investigate the Jacobson 2-radical structure of $PP(G)$ in order to see what the semisimple 
image is. 
An $N$-group $\Gamma$ is of \emph{type $2$} if there is $\gamma \in \Gamma$ such that $N\gamma = \Gamma$ and there do not exist non-trivial $N$-subgroups in $\Gamma$. 

\begin{defn}\cite{pilzbook}
 Let $N$ be a nearring. Then the \emph{Jacobson 2-radical} of $N$, written ${\cal J}_2(N)$, is the intersection of the 
 annihilators of all $N$-groups of type $2$.
\end{defn}

\section{Amnesia and Procrastination}

In this section we introduce two classes of automata and use them to bound ${\cal J}_2(PP(G))$.

\begin{de}
The map $\alpha:PP(G)\rightarrow PP(G)$ defined by: $\forall n \in PP(G)$, $\forall x \in G^\N$
\[((\alpha n)x)_i = (n(0,0,\ldots,0,x_i,0,\ldots))_i\]
is called the \emph{amnesiac map}.
\end{de}

\begin{lemma}
 The amnesiac map is a nearring homomorphism.
\end{lemma}
Proof:
Let $n,m \in PP(G)$, $x \in G^\N$.
Then
\begin{align}
 (\alpha (n+m)x)_i &= ((n+m)(0,\ldots,0,x_i,0,\ldots))_i \\
  &= (n(0,\ldots,0,x_i,0,\ldots) + m(0,\ldots,0,x_i,0,\ldots))_i \\
  &= (n(0,\ldots,0,x_i,0,\ldots))_i + (m(0,\ldots,0,x_i,0,\ldots))_i \\
  &= ((\alpha n)x + (\alpha m)x)_i
\end{align}
so $\alpha (n+m) = \alpha n + \alpha m$.

\begin{align}
 ((\alpha (nm))x)_i &= (nm(0,\ldots,0,x_i,0,\ldots))_i \\
  &= (n(m(0,\ldots,0,x_i,0,\ldots)))_i \\
  &= (n(0,\ldots,0,((\alpha m)x)_i,0,\ldots))_i \\
  &= ((\alpha n)((\alpha m) x))_i \\
  &= ((\alpha n)(\alpha m)x)_i
\end{align}
so $\alpha (nm) = (\alpha n)( \alpha m)$ and we see that $\alpha$ is a nearring homomorphism,.
\hfill$\Box$

\begin{lemma}
Let $n \in PP(G)$. The following are equivalent:
\begin{enumerate}
\item $n \in \ker \alpha$
\item $\forall x\in G^\N,\, ((\alpha n)x)_i = (n(0,0,\ldots,0,x_i,0,\ldots))_i =0$
\item $\forall i,\,n(\underbrace{0,0,\ldots ,0}_{i-1},x,\ldots) = (\underbrace{0,0,\ldots,0,0}_{i},\ldots)$
\item $\forall i,\,(n(\underbrace{0,0,\ldots ,0}_{i-1},x))\vert_{\{1,\ldots,i\}} = (0,0,\ldots,0) \in G^i$
\end{enumerate}
\end{lemma}
This can be seen because 2 is a rewording of the definition in 1, while 3 is simply rewriting 2. The fourth statement uses the restricted action to say the same as 3.

\begin{defn}
Let $f\in M_0(G)$ and $i,j \in \N$. 
Define $f^i \in PP(G)$ as $(f^i(x))_i = f(x_i)$ and $(f^i(x))_j = 0$ $\forall j\neq i$. 
Define $f^{i,j} \in PP(G)$ as $(f^{i,j}(x))_{i+kj} = f(x_{i+kj})$ for all $k\in \N_0$ and $(f^i(x))_l = 0$ otherwise. 
Let $M_{i,j}(G)= \{f^{i,j}: f \in M_0(G)\}$.
Then $M_{i,0}(G)= \{f^i: f \in M_0(G)\}$.
\end{defn}

The $M_{i,0}(G)$ automata react only at one time step and are elsewhere zero, ignoring any inputs before or after that time step. 
In some sense these are the most amnesiac of automata. Note that $\alpha f^i = f^i$ and $\alpha f^{i,j} = f^{i,j}$.

 \begin{lemma}
 Let $G$ be a group. Then
  \[J_2(PP(G)) \subseteq \ker \alpha\]
 \end{lemma}

 Proof: Suppose $n \not\in \ker \alpha$.
  $n \not\in \ker \alpha $ iff $\exists \bar g\in G $ such that $(n(0,\ldots,0,\bar g,\ldots))_i \neq 0$ with $\bar g$ in the $i$th place.
  
  Define an action of $PP(G)$ on $G$ as: for $m \in PP(G)$, $g \in G$, 
  $mg = (m(0,0,\ldots,0,g,\ldots))_i $ with $g$ in the $i$th place.  We need to show that $mg$ is indeed an action of $PP(G)$ on $G$.
  The additive property is clear, while the  multiplicative property is shown by
  \begin{align}
   (nm)g &= ((nm)(0,\ldots,0,g,\ldots))_i \\
     &= (n(m(,\ldots,0,g,\ldots))_i \\
     &= (n(0,\ldots,0,mg,\ldots))_i\\
     &= n(mg)
  \end{align}

  For all $f\in M_0(G)$, $f^i \in PP(G)$ acts as $f$ under this action, i.e.\ $f^ig=f(g)$. 
  
  So the action is the same as the action of $M_0(G)$ on $G$. Consequently, $G$ is a $PP(G)$ group of type $2$ which means that $J_2(PP(G)) \subseteq (0:G)$. But under this action, $n \not \in (0:G)$ because $n\bar g \neq 0$ , so $n \not \in  J_2(PP(G))$.
\hfill$\Box$

A procrastinating or delaying automaton never reacts immediately to an input. 
Maybe it should be called a burocrat?

 \begin{de}
  $n \in PP(G)$ is \emph{delaying} if 
  \[\forall k \in \N: x_i=y_i \forall i < k \Rightarrow (nx)_i=(ny)_i \forall i \leq k\]
 \end{de}
 
 We write $D(G)$ for this set of state machines.
 \begin{lemma}
   $D(G)$ is an $N$-subgroup.
 \end{lemma}
Proof: 
By rudimentary calculations it is clear that $D(G)$ is closed under addition and composition, forming a subnearring.

Now we need to show $PP(G)D(G) \subseteq D(G)$. 
Let $n\in PP(G)$, $d\in D(G)$, $x,y \in G^\N$ with $x_i=y_i$ for $i<k$.
Then $(dx)_i = (dy)_i$ for $i \leq k$ by the delaying property, so $((nd)x)_i=(n(dx))_i = (n(dy))_i=((nd)y)_i$ for $i \leq k$ by the prefix preserving property, so
$nd \in D(G)$.

Thus we see that $D(G)$ is an $N$-subgroup.
\hfill$\Box$

The following result gives us an idea of what the automata in $D(G)$ are like.
 \begin{prop}
  $n\in D(G)$ iff $n$ is a Moore machine iff every state output map is constant. 
 \end{prop}
Proof:
Being a Moore machine is defined as having every output map being constant.
So we need only concern ourselves with the first and last statements.

$(\Rightarrow)$: Let $n = (Q,t,f,s) \in D(G)$, so for all reachable $r \in Q$, there is some input sequence $x \in G^\N$ such that $r=q_{i+1}$ in the state sequence for some $i$.
Then by the delaying property, for all $g_1,g_2 \in G$,  $f(r,g_1)=n(x_1,\ldots,x_i,g_1,\ldots)_{i+1} = n(x_1,\ldots,x_i,g_2,\ldots)_{i+1}=f(r,g_2)$ so $f_r: g \mapsto f(r,g)$ is constant.
Thus every state output map is constant.

$(\Leftarrow)$:  Let $n = (Q,t,f,s) \in PP(G)$ with $f_q: g \mapsto f(q,g)$  constant for all $q \in Q$.
Then for all $x,y\in G^\N$ with $x_i=y_i$ $\forall i <k$, the state sequences induced by $x$ and $y$ will match up to $q_k = t(q_{k-1},x_{k-1})$.
Then $(nx)_k = f(q_k,x_k) = f(q_k,y_k) = (ny)_k$ so $n \in D(G)$ and we are done.
\hfill$\Box$

 \begin{theorem} Let $G$ be a group. Then
  $D(G) \leq J_2(PP(G)) \leq \ker \alpha$.
 \end{theorem}

 Proof: We know the second inclusion from above. So we need only show the first inclusion.
 
 Let $n\in D(G)$, as a state machine $(Q,t,f,s)$. 
 The automaton $1-n$ can be written as $(Q,t,h,s)$ with $h(q,g) = g-f(q,g)$.
 As every state output map $f_q:g \mapsto f(q,g)$ is a constant map by the claim above, the automaton $1-n$
 has output map $h(q,g)=g-f(q,g)$, state output maps $h_q:g \mapsto g-f_q(g)$ that are permutations. 
 Let $\bar h_q:G \rightarrow G$ be the inverse of this permutation, $\bar h(q,g) = \bar h_q(g)$.
 Define $\bar t:Q\times G \rightarrow Q$ by $\bar t(q,g) = t(q,\bar h(q,g))$.
 Then the state machine $m=(Q,\bar t, \bar h, s)$ is the inverse of $1-n$.
 
 The composition state machine $m \circ (1-n) = (Q\times Q, T, F, (s,s))$ for some maps $T$ and $F$. Note that the start state is on the diagonal. 
 Let $q \in Q,\,g\in G$. Then
 \[
  T((q,q),g) = (\bar t(q,h_q(g)), t(q,g)) =  (t(q,\bar h_q(h_q(g))),t(q,g)) = (t(q,g),t(q,g))
 \]
 so we see that the state remains on the diagonal $\{(q,q):q\in Q\} \subset Q \times Q$.
 Thus we need only consider the output maps on the diagonal, as no other states are reachable.
 The output function of the composition can be seen by
 \[
  F((q,q),g) = \bar h(q,h(q,g)) = \bar h_q (h_q(g)) = g
 \]
to be  the identity, $m \circ (1-n) = 1$, so by Beidleman  \cite[3.37c]{pilzbook}, $n$ is quasiregular. 

Thus the whole $N$-subgroup $D(G)$ is quasiregular, so by Ramakotaiah \cite{ramakotaiah67} (also \cite[Theorem 5.44]{pilzbook}) $D(G)$ lies within the
Jacobson 2-radical ${\cal J}_2(PP(G))$ and we are done.
 \hfill$\Box$
 
We know that the radical is an ideal, so we are interested in the ideal generated by $D(G)$. Our job is to work out when these nearrings coincide as ideals.

\section{${\cal J}_2$ for certain groups}

In this section we show that there are a class of groups where we can identify the Jacobson 2-radical of the prefix preserving maps on that group.

\begin{defn}
\label{defnX}
A group $(G,+)$ has \emph{property X} if there is an element $k\in G$ and a function $f:G \rightarrow G$ such that
$f(x+k)-f(x)=x$ for all $x \in G$.
\end{defn}

\begin{theorem}
 Finite abelian groups  have property X iff they are of odd order.
\end{theorem}

Proof: First we show that all cyclic groups of odd order have property X. Then we show that odd order abelian groups have property X. Then we will show that a group with an element of even order will not have property X.

Let $n>1$ be an odd integer. Let $k=1 \in \Z_n$ be a generator. Let $x_i=f(i)$.
Property X can be written as a collection of linear equations over $Z_n$ with $-x_i + x_{i+1} = i$ for $0\leq i <n$ with suffix addition modulo $n$. 
The $n$th equation is $-x_{n-1} + x_0 = n-1$.
The sum of the first $n-1$ equations is
\[
 -x_0 +x_1 - x_1 + x_2+\ldots +x_{n-1} = 0+1+2+\ldots+{n-2}
\]
which reduces to $-x_0 + x_{n-1} = 1$, the additive inverse of the $n$th equation. 
Thus the $n$th equation is redundant and we are left with $n-1$ equations having a set of solutions with one parameter. 
We select $x_0$ then for each $0<i<n$,  $x_i=x_{i-1}+i-1$.

Let $G$ be an abelian group of odd order. Let $H$ be a maximal cyclic subgroup of $G$, generated by $k$ of order $2n+1$. 
$H$ has property X by the above argument. 
Let $c \not\in H$. Then in $c+H$ the same argument holds (that the last equation is redundant) because the order of $c$ divides $2n+1$ and thus $(2n)c = -c$. 
Thus in each coset of $H$ we can find values for $f$ and we are done.

Suppose $G$ is an abelian group of even order satisfying property X. 
If the element $k$ has even order, say $n=2m$, then the set of  the first $n-1$ equations, when added, gives $x_0 - x_{n-1} = 1+m$.
The $n$th equation is $x_{n-1} - x_0 = n-1$ so we add them to get $0=n-1+1+m = n+m$, i.e.\  $n=m=0$, a contradiction.
Thus the element $k$ for property X should be of odd order, let the order be $n$. Let $c\in G$ be of order 2. Then there will be a set of $n$ equations of the form
\[
f(c+ik) - f(c+(i+1)k) = c+ik
\]
Adding all these equations together we get
\[
0 = nc + \frac{n(n+1)}{2}k = nc
\]
but $c$ is of even order and $n$ is odd, a contradiction.

Thus no abelian group of even order can have property X.
\hfill$\Box$

\begin{lemma}
Infinite cyclic groups have property X. 
\end{lemma}
Proof: Suppose $G$ is an infinite order cyclic group, we will assume it is $\Z$. 
Defining $f(0)=f(1)=\ldots=f(k-1)=0$ and $f(x+k)=f(x)+x$ otherwise gives a function for property X.
\hfill$\Box$

Note that without loss of generality, we can insist that $f(0)=0$.


\begin{theorem}
 Let $(G,+)$ be a group with property X. Then ${\cal J}_2(PP(G)) = \ker \alpha$.
\end{theorem}
Proof:
Let $d\in \ker \alpha$. 

Let $g\in G$ be some arbitrary but fixed nonzero element.
Let $c\in D(G)$ be the  delaying automaton defined as $(\{a,b\},t,o,a)$
with $t(a,0)=a$, $t(a,x)=t(b,y)=b$ and
 $o(a,y)=0$, $o(b,y)=g$ for all $x,y \in G, x\neq 0$.

  The initial $0$ inputs are mapped to $0$, the first nonzero input as well, then all outputs are $g\neq 0$.

Let $f$ be a one state automaton with the output function being the function defined by property X.
Then $f(d+c) - fd = d$ so the left ideal generated by $D(G)$ is all of $\ker \alpha$ which is an ideal. 
Thus ${\cal J}_2(PP(G)) = \ker \alpha$.
\hfill$\Box$

\section{Semisimplicity}

The radical is, in some sense, the ``bad'' part of a nearring. So we factor that out to get the ``good'' part. 
What is the image of $PP(G)$ under the map $\alpha$?

The following result shows what the amnesiac map does to an automaton.

\begin{lemma}
Let $n=(Q,t,o,s)$ be an automaton.
Define  $\bar t (q,g) = t(q,0)$ $ \bar n = (Q,\bar t,o,s)$.
Then $\bar n = \alpha n$ and $\alpha$ is idempotent.
\end{lemma}
Proof: 
For any $i$, 
$(\alpha n x)_i = (n(0,\ldots,0,x_i,0,\ldots))_i$, the state sequence $q$ induced by $(0,\ldots,0,x_i,0,\ldots)$ in $\alpha n$
agrees, up to the $i$th entry, with the state sequence induced by $x$ in $\bar n$.
So $(\alpha n x)_i = o(q_i,x_i) = (\bar n x)_i$.
Thus $\alpha n$ and $\bar n$ agree for all $i$, so they are equal.

$\bar \bar t (q,g) \bar t(q,0) = t(q,0)= \bar t(q,g)$, so $\bar \bar n = \bar n$ so the mapping $\alpha$ is idempotent.
\hfill $\Box$

\begin{theorem} 
Let $(G,+)$ be a group such that ${\cal J}_2(PP(G)) = \ker \alpha$. 
Then, $PP(G) / {\cal J}_2(PP(G)) = \sum M_{i,0}(G)$.
\end{theorem}
Proof:
First we show that the sum is direct.
Let $i\neq j$, $f^i \in M_{i,0}(G),\, g^j \in M_{j,0}(G)$.
Then $(f^ig^j(x))_i = f(g^j(x)_i) = f(0) = 0$, 
 and for $k \neq i$, $(f^ig^j(x))_k =  0$. So $f^ig^j=0$ and we are done.

Let $n = (Q,t,f,s)$ be an automaton, $\alpha n = n$. Then $t(q,g)=t(q,0)$ for all $q\in Q$.
Define $\tau :Q\rightarrow Q$ by $\tau: q \mapsto t(q,0)$.
Then the state sequence will be $q_i = \tau^{i-1}s$.

Define $f_i(g) = f(\tau^{i-1}s,g) \in M_0(G)$.
Then we claim that
\[n = \sum_{i\in \N} f_i^{i,0}
\]

Thus $\alpha n = n \Rightarrow n \in \sum M_{i,0} (G)$.
We know the converse, so  $\alpha PP(G) = \sum M_{i,0}(G)$.

Now $PP(G) / {\cal J}_2(PP(G)) = PP(G) / \ker \alpha  = \alpha(PP(G))$, so we are done.
\hfill$\Box$

We note that a finite state automata will give, in general, an infinite sum in $\sum M_{i,0}(G)$.
It will, however, give a finite sum in $\sum M_{i,j}(G)$, by summing over the 0-reachable states.

\section{Summary}

We examined the properties of nearrings of state automata and found that the nearrings are highly complex.
In order to get a handle on them, we  looked at the radicals and the semisimple images modulo these radicals.
The Jacobson 2-radical could be constrained and, in the case of noneven order abelian groups, fully determined.
It equals the kernel of the amnesiac endomorphism.
Thus we can say that the representatives of the 2-semisimple nearring of automata are the amnesiac automata. 
These are sequences of mappings from $M_0(G)$. 
These sequences are infinite, but in the case of finite state automata they will cycle after an initial sequence.

 In difference to ring theory, several different types of simplicity of $N$-groups
exist (see \cite{pilzbook}).
One obvious next stage would be to determine other radicals in $PP(G)$ and to see in which way their semisimple images behave as automata.

In difference to ring theory the radical theory for nearrings is much
  more complex. Several
  types of simplicity for $N$-groups of a nearring $N$ exist. Apart from $N$-groups of type $2$ the most common ones are $N$ groups of type $0$ and 
  type $1$. The intersection of the annihilators of the $N$-groups of type $1$, type $0$ respectively, defines the 
  \emph{Jacobson 1-radical} ${\cal J}_1(N)$, \emph{Jacobson 0-radical} respectively. Note that $PP(G)$ contains the identity 
  mapping, so  ${\cal J}_2(N)={\cal J}_1(N)$ by \cite[Proposition 5.3]{pilzbook}. There are also a range of further radicals, all of which would offer further insights.

On the other hand, it
would be of interest to determine the 2-radical for further classes of groups.

\bibliographystyle{plain}

\bibliography{automataradicals}

\section*{Acknowledgments}

The authors gratefully acknowledge support from the Austrian Science Fund FWF under project number 23689-N18.

\end{document}